# Towards an Exact Mechanical Analogy of Particles and Fields.

**Valery P. Dmitriyev**

Lomonosov University, Moscow, Russia
P.O. Box 160, Moscow 117574, Russia;
e-mail: dmitr@cc.nifhi.ac.ru

**Abstract**

An exact analogy of electromagnetic fields and particles can be found in continuum mechanics of a turbulent perfect fluid with voids. Deviations of the turbulence from a homogeneous isotropic state correspond to electromagnetic fields: with the average pressure as electrostatic potential, the average fluid velocity as magnetic vector potential and the density of the average turbulence energy as electromotive force. The waves of turbulence perturbation model the electromagnetic waves. Cavities of the fluid serve as walls to support stationary perturbations of turbulence. Cavitation of the turbulent noncorpuscular fluid occurring in the presence of voids leads to forming dilatational inclusions of empty space and of the quiescent fluid. These model the positive and negative electrically charged particles, respectively. Due to the dilatation, the inclusions interact with the turbulence perturbation fields. This looks exactly as interaction of the charges with the electromagnetic fields. Splitting and dispersion of an inclusion in the stochastic environment model delocalization of a quantum particle.

## 1  Introduction

We are in search of mechanical medium capable to reproduce or imitate electromagnetics including charge and charged particles. Averaged turbulence in an inviscid incompressible fluid is considered. Following standard Reynolds scheme, an infinite chain of nonlinear equations for growing number of unknowns can be obtained. We choose among a variety of approximations the simplest one – minimal closure of the chain and linearizing the model. The consistent system of linear equations thus formed was found [1] to be isomorphic to the field part of Maxwell's electromagnetic equations.

The charge (and particle) portion of the theory will be shown below to be associated with voids in the fluid. The essential point is that the fluid is noncorpuscular. Hence, there is no entropy, no temperature and vapor phase can't be existent. These give rise to a whole spectrum of structures which reproduce the world of particles as sources of stationary fields.

## 2  Turbulence averaged

We describe dynamics of a fluid in terms of the flow velocity $\mathbf{u}(\mathbf{x},t)$ and the specific pressure $p(\mathbf{x},t)$. Following well known in hydrodynamics Reynolds technique, we consider (short-time temporal or statistical) averages of the velocity and pressure, $\langle \mathbf{u} \rangle$ and $\langle p \rangle$, which are also the functions of the space $\mathbf{x}$ and time $t$ coordinates. Whence the turbulent pulsations $\mathbf{u}'$ and $p'$ can be defined:
$$\mathbf{u} = \langle \mathbf{u} \rangle + \mathbf{u}',$$
$$p = \langle p \rangle + p'. \qquad (2.1)$$
The fluid is supposed to be incompressible, at least in its fluctuation component:
$$\partial_i u_i' = 0.$$
Putting (2.1) in Euler equation
$$\partial_t u_i + u_k \partial_k u_i + \partial_i p = 0, \qquad (2.2)$$
averaging and taking account of $\langle \mathbf{u}' \rangle = 0, \langle p' \rangle = 0$, we find for turbulence averages:
$$\partial_t \langle u_i \rangle + \langle u_k \rangle \partial_k \langle u_i \rangle + \partial_k \langle u_i' u_k' \rangle + \partial_i \langle p \rangle = 0. \qquad (2.3)$$
Here and further on $\partial_t = \partial / \partial t$, $\partial_k = \partial / \partial x_k$, $i,k=1,2,3$ and summation over recurrent index is implied throughout.

In the ground state and also at infinity the turbulence is supposed to be homogeneous and isotropic:
$$\langle \mathbf{u} \rangle^0 = 0,$$
$$\langle p \rangle^0 = \text{const}, \qquad (2.4)$$
$$\langle u_i' u_k' \rangle^0 = c^2 \delta_{ik}.$$
Integrating Reynolds equation (2.3) for the case of $\langle \mathbf{u} \rangle = \text{const}$, we may get
$$\langle u_i' u_k' \rangle = \langle u_1' u_1' \rangle \delta_{ik},$$
$$\langle u_1' u_1' \rangle + \langle p \rangle = c^2 + \langle p \rangle^0. \qquad (2.5)$$

This is a kind of Bernoulli equation and actually an equation of state of ideal isotropic turbulence. It implies rather a broad range of variation for the turbulence energy density $1/2\langle u'_i u'_i \rangle$ and pressure, involving coexistence of different turbulence phases.

## 3 Perturbations of turbulence

Equation (2.3) represents the first link in the chain of dynamical equations for consecutive moments of turbulent pulsations. Next equation is received multiplying (2.2) by $u'_l$, symmetrizing, substituting in it (2.1) and averaging:

$$\langle u'_k \left[ \partial_t (\langle u_i \rangle + u'_i) + (\langle u_j \rangle + u'_j) \partial_j (\langle u_i \rangle + u'_i) + \partial_i (\langle p \rangle + p') \right] + \\ + u'_i \left[ \partial_t (\langle u_k \rangle + u'_k) + (\langle u_j \rangle + u'_j) \partial_j (\langle u_k \rangle + u'_k) + \partial_k (\langle p \rangle + p') \right] \rangle = 0. \tag{3.1}$$

Consider small deviations from (2.4):

$$d\langle \mathbf{u} \rangle = \langle \mathbf{u} \rangle - \langle \mathbf{u} \rangle^0,$$
$$d\langle p \rangle = \langle p \rangle - \langle p \rangle^0,$$
$$d\langle u'_i u'_k \rangle = \langle u'_i u'_k \rangle - \langle u'_i u'_k \rangle^0.$$

Substituting them in (2.3), (3.1) and neglecting quadratic terms, we come [1] to linearized Reynolds equations, respectively:

$$\partial_t d\langle u_i \rangle + \partial_k d\langle u'_i u'_k \rangle + \partial_i d\langle p \rangle = 0, \tag{3.2}$$

$$\partial_t d\langle u'_i u'_k \rangle + c^2 (\partial_i d\langle u_k \rangle + \partial_k d\langle u_i \rangle) + h_{ik} = 0, \tag{3.3}$$

where

$$h_{ik} = \langle u'_i \partial_k p' \rangle + \langle u'_k \partial_i p' \rangle + \partial_j \langle u'_i u'_j u'_k \rangle.$$

Here $c$ takes the sense of the speed for the wave of turbulence perturbation propagating in the medium.

## 4 Maxwell's equations

Differentiating (3.3) with respect to $x_k$, we get vector equation

$$\partial_t \partial_k d\langle \mathbf{u}' u'_k \rangle - c^2 \nabla \times \nabla \times d\langle \mathbf{u} \rangle + \mathbf{g} = 0, \tag{4.1}$$

where

$$g_i = \partial_k h_{ik} + 2c^2 \partial_i \nabla \cdot d\langle \mathbf{u} \rangle$$

and the identity $\nabla(\nabla\cdot)=\nabla\times\nabla\times+\nabla^2$ was taken advantage of. In the points of incompressibility

$$\nabla\cdot d\langle\mathbf{u}\rangle=0. \qquad (4.2)$$

Next, following [1], we define vector and scalar quantities:

$$A_i = kc\, d\langle u_i\rangle, \qquad (4.3)$$

$$E_i = k\partial_k d\langle u'_i u'_k\rangle, \qquad (4.4)$$

$$j = k d\langle p\rangle, \qquad (4.5)$$

$$j_i = \frac{k}{4\pi}\partial_k h_{ik}, \qquad (4.6)$$

where $k$ is an arbitrary constant. Substituting (4.3)-(4.6) in (3.2) and (4.1), there are obtained two of Maxwell's equations, respectively,

$$\frac{1}{c}\frac{\partial \mathbf{A}}{\partial t}+\mathbf{E}+\nabla j=0,$$

$$\frac{1}{c}\frac{\partial \mathbf{E}}{\partial t}-\nabla\times(\nabla\times\mathbf{A})+\frac{4\pi}{c}\mathbf{j}=0,$$

with (4.3) in (4.2) giving the Coulomb gauge

$$\nabla\cdot\mathbf{A}=0.$$

The missing part of the theory is the electrical charge and interaction of the charges. As will be shown below, the charge and particles correspond to discontinuities of the medium.

Thus, it looks like mapping of conventional electromagnetics into minimal closure linear Reynolds turbulence: in (3.3) there must be $h_{ik}=0$ when the medium is continuous.

## 5 Cavitons

Next, we consider an incompressible noncorpuscular fluid with voids. A hollow bubble can not be in equilibrium with the ground state (2.4) of a turbulent fluid since $\langle p\rangle^0 \neq 0$. In molecular fluids the bubble will fill with the fluid vapor until the gas pressure becomes $\langle p\rangle^0$. However, here we deal with a true continuum. It does not consist of corpuscles and hence the vapor phase can not be formed. So, the empty bubble attains equilibrium via a perturbation of the fluid turbulence (2.4). This may proceed in the following way.

Let there be two hollow bubbles in the homogeneously turbulent incompressible fluid. The relative sizes of the bubbles are chosen with a view to simplify the later events. The system passes spontaneously into a stable configuration – the fluid cavitates by expanding one bubble and contracting the other, a redistribution of the turbulence energy between the bubbles thereby taking place. The bubble expanded is stabilized via dropping the local pressure $\langle p \rangle$ down to zero. By (2.5), this is accompanied by respective rise in the energy density (Fig.1, top). The other bubble shrinks in forming a stable islet of quiescent fluid i.e. the local energy density $1/2\langle u'_i u'_i \rangle$ decreases to zero and by (2.5) corresponding rise in the pressure takes place (Fig.1, bottom). This process must conserve the total turbulence energy

$$1/2 \int \mathbf{V} \langle u'_i u'_i \rangle d^3 x, \tag{5.1}$$

where $\mathbf{V}$ is the fluid density and $d^3x = dx_1 dx_2 dx_3$. Further, the centers, or sources, of medium perturbation thus formed will be referred to as *cavitons*.

The fluid displacement $\mathbf{s}$ produced by a caviton located at $\mathbf{x}'$ is found from the point model of dilatational inclusion

$$\nabla \cdot \mathbf{s} = \Delta V\, \delta(\mathbf{x} - \mathbf{x}'), \tag{5.2}$$

where $\mathbf{s}(\mathbf{x} - \mathbf{x}')$ is taken in Euler coordinates. A bubble expanded corresponds to $\Delta V > 0$, a bubble contracted – to $\Delta V < 0$. The simultaneous contraction of voids in one place and their expansion in the other means

$$\Delta V_1 + \Delta V_2 = 0 \tag{5.3}$$

that is consistent with microscopic incompressibility of the fluid.
Thus, a caviton combines in itself a center of medium dilatation with a center of turbulence perturbation.

## 6 Fields

In order to determine the shape of stationary perturbation fields produced by cavitation, we consider small oscillations in the system (5.3) of two opposite cavitons located a large distance from each other. In this event, $\Delta V$ is taken as a function of time $t$. For small perturbation values $\boldsymbol{d}\langle \mathbf{u} \rangle = \langle \mathbf{u} \rangle$ of the flow velocity, the quadratic terms in (2.3) can be neglected:

$$\partial_t \boldsymbol{d}\langle u_i \rangle + \partial_k \langle u'_i u'_k \rangle + \partial_i \langle p \rangle = 0. \tag{6.1}$$

Due to instantaneity of action peculiar to incompressible medium, the space and time variables in the fields can be separated. We have nearby a caviton:

$$\langle p \rangle = -[a(t) + \boldsymbol{d}a(t)]\, f(\mathbf{x} - \mathbf{x}') + \langle p \rangle^0,$$

$$\langle u'_i u'_k \rangle = \left[ a(t) f(\mathbf{x} - \mathbf{x}') + c^2 \right] \delta_{ik},$$

where $f(r) \to 0$ when $r \to \infty$. Assuming for this kind of the motion

$$d\langle \mathbf{u}\rangle = \partial_t \mathbf{s}$$

and operating on (6.1) with $\partial_i$ gives through (5.2)

$$d(\mathbf{x}-\mathbf{x}')\P_t^2 \Delta V(t) = \nabla^2 f(\mathbf{x}-\mathbf{x}') da(t).$$

Resolving equation

$$\nabla^2 f \sim -\delta(\mathbf{x}-\mathbf{x}'),$$

we get

$$f(\mathbf{x}-\mathbf{x}') \sim 1/|\mathbf{x}-\mathbf{x}'|. \tag{6.2}$$

Taking account of the boundary conditions on the walls of the cavitons, we have for a stable hollow bubble of the radius $R$

$$\langle p \rangle = -\frac{\langle p \rangle^0 R}{|\mathbf{x}-\mathbf{x}'|} + \langle p \rangle^0, \tag{6.3}$$

$$\langle u_i' u_k' \rangle = \left( \frac{\langle p \rangle^0 R}{|\mathbf{x}-\mathbf{x}'|} + c^2 \right) \delta_{ik}, \tag{6.4}$$

(Fig.1, top), and for an islet of quiescent fluid of the radius $r_e$

$$\langle u_i' u_k' \rangle = \left( -\frac{c^2 r_e}{|\mathbf{x}-\mathbf{x}'|} + c^2 \right) \delta_{ik}, \tag{6.5}$$

$$\langle p \rangle = \frac{c^2 r_e}{|\mathbf{x}-\mathbf{x}'|} + \langle p \rangle^0, \tag{6.6}$$

(Fig.1, bottom), where (2.5) was used. Because of conservation of the turbulence energy (5.1), there must be in this relations

$$\langle p \rangle^0 R = c^2 r_e = |a|, \tag{6.7}$$

or

$$r_e / R = \langle p \rangle^0 / c^2.$$

The form (6.2) of perturbation fields thus obtained should be corrected near the core of the cavitons, since we have omitted (see (6.1)) in Reynolds equation (2.3) the convective term $\langle u_k \rangle \partial_k \langle u_i \rangle$. In a nonlinear model, a dilatational inclusion can be associated with a vortex, e.g. a vorton, that attaches to itself an additional stability and spin properties.

## 7  Static interaction

It is known that spherical dilatational inclusions of a linear elastic medium don't interact distantly with each other or with a stress field. However, spherical dilatational inclusions of a turbulent fluid do interact, since they generate in the medium both displacement and pressure fields. The work $\boldsymbol{e}$ done against a perturbation pressure field $\boldsymbol{d}\langle p \rangle = \langle p \rangle - \langle p \rangle^0$ while creating an inclusion (5.2) is given by

$$\boldsymbol{e} = -V \int \boldsymbol{d}\langle p \rangle \nabla \cdot \mathbf{s} \, d^3 x = -V \Delta V \boldsymbol{d}\langle p \rangle. \qquad (7.1)$$

Taking account of (6.3), (6.6) and (6.7), that gives for two cavitons separated by distance $\left| \mathbf{x}^{(1)} - \mathbf{x}^{(2)} \right|$:

$$\boldsymbol{e}_{12} = 1/2 V (\Delta V_1 a_2 + \Delta V_2 a_1) / \left| \mathbf{x}^{(1)} - \mathbf{x}^{(2)} \right|, \qquad (7.2)$$

where $\text{sign}(a) = \text{sign}(\Delta V)$. So, two cavitons with opposite signs of $\Delta V_1$, $\Delta V_2$ attract each other, while those with like signs of $\Delta V_1$, $\Delta V_2$ repel each other, behavior resembling Coulomb law from electrostatics.

## 8  Electrical charge and elementary particles

Defining electrical charge $q$ by

$$q = -V \Delta V / \boldsymbol{k}, \qquad (8.1)$$

where $\boldsymbol{k} < 0$, and using (4.5), we find in (7.1) a mechanical analogue for the energy $q\boldsymbol{j}$ of electrostatic interaction.

However, the mechanical analogy appears to be incomplete, since reciprocity of interaction (Betti's theorem) does not hold in this medium. So, we have expression (7.2) instead of the form

$$\boldsymbol{k}^{-2} V^2 \Delta V_1 \Delta V_2 / \left| \mathbf{x}^{(1)} - \mathbf{x}^{(2)} \right|$$

needed to reproduce Coulomb's law. Obviously, the latter is obtained from (7.2) only when

$$a = V \Delta V / \boldsymbol{k}^2 \qquad (8.2)$$

On the other side, from (6.4), (6.5) and (6.7) we have the relation

$$\partial_i \partial_k \langle u_i' u_k' \rangle = -4 \boldsymbol{p} a \, \delta(\mathbf{x} - \mathbf{x}'). \qquad (8.3)$$

With account of (4.4), it reproduces the formula of electromagnetism

$$\nabla \cdot \mathbf{E} = 4 \boldsymbol{p} q^* \, \delta(\mathbf{x} - \mathbf{x}'),$$

however, providing that a different definition of electrical charge $q^*$ is assumed:

$$q^* = -\boldsymbol{k} a. \qquad (8.4)$$

Postulating identity of the field (8.4) and particle (8.1) definitions of charge

$$q^* = q$$

returns us to relation (8.2).

Taking account of (6.7), that actually means a unique value for the strength $a$ and volumes

$$\frac{4\boldsymbol{p}}{3} R^3 = V, \qquad \frac{4\boldsymbol{p}}{3} r_e^3 = |\Delta V|$$

of the cavitational inclusions. This can be the case if only the turbulent medium has a discrete microscopic structure. For instance, in the "vortex sponge" [2] $R$, or $r_e = R \langle p \rangle^0 / c^2$, must be determined by the fundamental constant $L$ of the structure – the vortex filament length per unit volume of the medium. Thus, we come to discrete units of matter and charge i.e. to elementary particles. In this event, the hollow bubble shown in Fig.1, top corresponds to the proton, the islet of quiescent fluid shown in Fig.1, bottom – to the electron. The neutron is modeled by a non-equilibrium hollow bubble of a homogeneously turbulent fluid. The masses of the particles are measured by the equivalent mass of the fluid:

$$M=W, \qquad m_e=V|\Delta V|.$$

## 9 Physical vacuum

In order to construct properly the electron and proton, we need for relative sizes of the two opposite cavitons to be

$$r_e \ll R.$$

According to (6.7), this is provided by

$$\langle p \rangle^0 \ll c^2. \qquad (9.1)$$

On the other side, in an ideal fluid

$$p>0.$$

Hence, in oscillatory waves, disturbances are restricted by

$$|d\langle p \rangle| < \langle p \rangle^0.$$

Combining this with (9.1) and using (2.5), we come automatically to requirements of the linearized model:

$$|d\langle u'_i u'_k \rangle|, |d\langle p \rangle|, d\langle u \rangle^2 \ll c^2. \qquad (9.2)$$

So, one may conclude that physical vacuum should be represented by the high-energy low-pressure turbulence (9.1) in a perfect fluid.

## 10 Distinction between vacuum and electromagnetic energies

Following section 8, let us combine (8.3) with (5.2):

$$a\nabla \cdot \mathbf{s} = -\Delta V \partial_i \partial_k \langle u'_i u'_k \rangle. \qquad (10.1)$$

That gives from (3.2)

$$a\nabla\cdot\mathbf{s}=\Delta V\nabla^2 d\langle p\rangle. \qquad (10.2)$$

This is a kind of polylinear elasticity. The equation of state (10.2) is quite distinct from the Hooke's law, even if $\Delta V/a$ is fixed. Then, starting from the substitution of (10.1) to the integral in (7.1), various forms or expressions of the elastic energy can be obtained. These are the familiar formulas, which provide us with mechanical analogies for different parts of electromagnetic energy.

Thus, two kinds of the energy should be distinguished: the turbulence energy (5.1) and the elastic energy of the turbulent medium. These can be interpreted as vacuum and electromagnetic energies, respectively. Really, they do not intermix with each other directly. For example, we have for the plane wave in the continuous medium:

$$h_{ik}=0, \qquad d\langle p\rangle=0,$$

$$d\langle\mathbf{u}\rangle=\mathbf{l}F(wt-\mathbf{k}\cdot\mathbf{x}), \qquad \mathbf{k}\cdot\mathbf{l}=0,$$

$$d\langle u'_i u'_j\rangle=\frac{c^2}{w}\left(l_i k_j+l_j k_i\right)F(wt-\mathbf{k}\cdot\mathbf{x}).$$

Here the density of the perturbation turbulence energy is vanishing

$$d\langle u'_i u'_i\rangle=\frac{2c^2}{w}k_i l_i F=0,$$

while the terms of the elastic energy $\left(\partial_k\langle\mathbf{u}'u'_k\rangle\right)^2$ and $\left(\nabla\times d\langle\mathbf{u}\rangle\right)^2$ are different from zero.

## 11  Dispersing a caviton

We assume that in the continuous fluid $h_{ik}=0$. Taking in (3.3) $k=i$ and summing over the recurrent index gives for the incompressible medium

$$\partial_t d\langle u'_i u'_i\rangle=0. \qquad (11.1)$$

So, in the continuous incompressible fluid, the profile of the turbulence energy remains unchanged in the course of evolution of the perturbations (9.2).

However, in a discontinuous fluid, relation (11.1) does not hold. A source (or center) of perturbation can be smeared up and distributed over the medium under the action of turbulent pulsations, the total turbulence energy (5.1) and $\Delta V$ being conserved. The profile of the turbulence perturbation in a distributed caviton may conform to the condition (9.2) of the linearized model. While in the localized negative caviton it does not (Fig.1, bottom). So, the half-width of the core of the distribution shown in Fig.2 can be taken as the "classical radius of the electron" $R_e$. It is easily found from conservation of the turbulence energy (5.1) and the geometry of the distribution:

$$R_e \approx 1.5 R.$$

In general, the volume distribution of a caviton can be viewed as a dispersion of splinters of the caviton. It is conveniently expressed in terms of the distribution density $c(\mathbf{x}',t)$ which is normalized by

$$\int c(\mathbf{x}',t) d^3 x' = \Delta V = \text{const}. \qquad (11.2)$$

In this event, the dilatation center (5.2) turns into a plasma of the point dilatation which is described by

$$\nabla \cdot \mathbf{s} = c(\mathbf{x},t).$$

Accordingly, we have instead of (8.3):

$$k \partial_i \partial_k \langle u'_i u'_k \rangle = -\frac{4\pi V}{k} c(\mathbf{x},t), \qquad (11.3)$$

where (8.2) was used. With account of the definitions (4.4), (8.1), that corresponds exactly to electro-magnetic relation for the charge density $r(\mathbf{x}',t)$:

$$\nabla \cdot \mathbf{E} = 4\pi r(\mathbf{x},t).$$

Via (11.3), (11.2), an equation of motion of the plasma of a point discontinuity closes the chain of equations (4.2), (3.2), (3.3). In this connection, it can be shown that $c$ evolves exactly as the density $\left|\Psi^*\Psi\right|$ of the wave function (see also [3]).

## 12  Conclusion

Summarizing, it is claimed that averaged turbulent medium provides an adequate theoretical framework for modeling classical electrodynamics. In the whole, we have the following correspondence between the terms of conventional electromagnetics and their mechanical counterparts.

| Electromagnetic concept | Property of a turbulent fluid |
|---|---|
| electromagnetic wave | turbulence perturbation wave |
| magnetic vector-potential | average velocity of the fluid flow |
| electrostatic potential | average fluid pressure |
| electromotive force | difference in the density of the average turbulence energy |
| formation of the pair $\left(p^+, e^-\right)$ of opposite electrical charges | turbulent cavitation of a discontinuous fluid |
| electrical current | flow of the point dilatation |
| mass of a particle | equivalent mass of a cavity in the fluid |

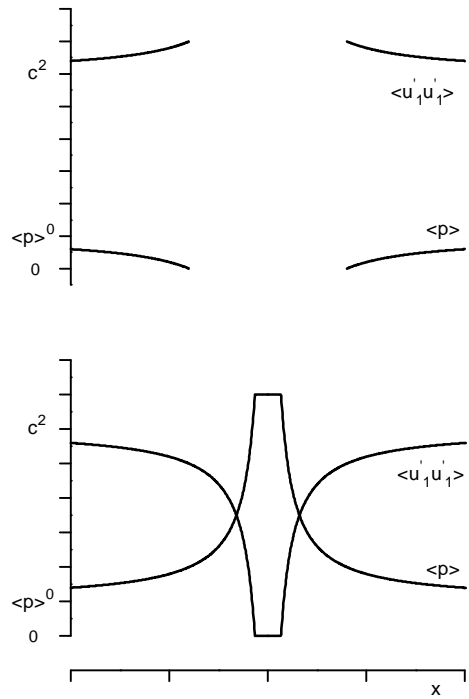

Fig. 1. – Positive (above) and negative (below) cavitons, representing the proton and the electron.

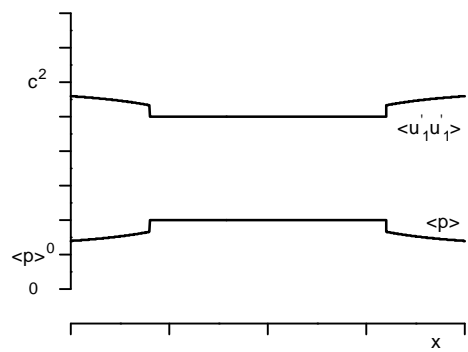

Fig. 2. – The negative caviton (Fig.1, bottom) flattened to the level of the wave amplitude.

See:

The earlier version of this work has been delivered in 1996 to the British Society for Philosophy of Science Meeting "Physical Interpretation of Relativity Theory".